\documentclass[a4paper,twoside,anonymous]{article}

\usepackage{epsfig}
\usepackage{subcaption}
\usepackage{calc}
\usepackage{amssymb}
\usepackage{amstext}
\usepackage{amsmath}
\usepackage{amsthm}
\usepackage{multicol}
\usepackage{pslatex}
\usepackage{apalike}
\usepackage{dirtytalk}
\usepackage{algorithmic}
\usepackage{graphicx}
\usepackage{makecell}
\usepackage{caption} 
\usepackage{textcomp}
\usepackage{xcolor}
\usepackage{hyperref}
\usepackage[hang,flushmargin]{footmisc}
\usepackage{SCITEPRESS}     

\begin{document}

\title{Labour Market Information Driven, Personalized, OER Recommendation System for Lifelong Learners}

\author{\authorname{Mohammadreza Tavakoli\sup{1}\orcidAuthor{0000-0002-7368-0794}, Stefan T Mol\sup{2}\orcidAuthor{0000-0002-9375-3516} and Gábor Kismihók\sup{1}\orcidAuthor{0000-0003-3758-5455}}
\affiliation{\sup{1}TIB Leibniz Information Centre for Science and Technology, Hannover, Germany}
\affiliation{\sup{2}University of Amsterdam, Amsterdam, Netherlands}
\email{\{reza.tavakoli, gabor.kismihok\}@tib.eu, s.t.mol@uva.nl}
}


\abstract{
 In this paper, we suggest a novel method to aid lifelong learners to access relevant OER based learning content to master skills demanded on the labour market. Our software prototype 1) applies Text Classification and Text Mining methods on vacancy announcements to decompose jobs into meaningful skills components, which lifelong learners should target; and 2) creates a hybrid OER Recommender System to suggest personalized learning content for learners to progress towards their skill targets. For the first evaluation of this prototype we focused on two job areas: \emph{Data Scientist}, and \emph{Mechanical Engineer}. We applied our skill extractor approach and provided OER recommendations for learners targeting these jobs. We conducted in-depth, semi-structured interviews with 12 subject matter experts to learn how our prototype performs in terms of its objectives, logic, and contribution to learning. More than 150 recommendations were generated, and 76.9\% of these recommendations were treated as useful by the interviewees. Interviews revealed that a personalized OER recommender system, based on skills demanded by labour market, has the potential to improve the learning experience of lifelong learners.
}

\onecolumn \maketitle \normalsize \setcounter{footnote}{0} \vfill

\section{\uppercase{Introduction}}
\label{sec:introduction}
The worlds of work and employment are changing rapidly in our post-industrial societies. As a consequence, matching processes between skill demand and supply are getting more and more complicated as skills dynamically evolve through an uncontrollable process \cite{colombo2018applying,castello2014promoting}. These dramatic changes lead to a number of educational problems in relation to the gap between (dynamic) skills that job markets demand and the training that education programs offer \cite{smith2014analyzing,wowczko2015skills,mcgill2009defining}. Furthermore, being up to date about actual job market skills has significant importance for individuals to remain employed or climb workplace hierarchy during active times of employment \cite{colombo2018applying,khobreh2015ontology}. Notably, in order to mitigate mismatches between education and labour markets, we need to 1) understand the dynamic nature of labour markets, which requires the deconstruction of jobs into required skills, and 2) match those skills to relevant learning content.

In order to tackle the first problem, governments and international organizations have created a number of occupational and skill taxonomies to provide structure for job-seekers and employers about skill components of jobs (e.g. ESCO, ISCO, O*NET). However, there are obstacles limiting the usefulness of these taxonomies, such as keeping their information updated \cite{djumalieva2018open}. At the same time, researchers attempt to build ontologies to provide accurate representations of jobs and skills (e.g. \cite{sibarani2017ontology}), and machine learning models to capture information from rich, text based, labour market data sources, like job vacancy announcements \cite{colace2019towards,boselli2018classifying,boselli2018wolmis,kobayashi2018text}.

To address the second issue, educational services should be tailored to the needs of individual lifelong learners. In this respect, open education become a key facilitator in many areas, including personal skills development \cite{kanwar2018global}. Open Educational Resources (OERs) are also gaining popularity as content sources for open education \cite{ha2011novel}. Major OER repositories have large amount of regularly updated learning content in wide range of content areas. Therefore it is surprising that despite their growing capacity, OER platforms still under perform when offering personalised learning services. As an example, OER users must consult and search through several OER repositories (with different interfaces) manually in order to find appropriate learning content. Only few, initial efforts are reported, which attempt to build OER recommendation algorithms. These are done by collecting properties of users and OERs using various approaches such as building (or reusing existing) ontologies \cite{wan2018learning,sun2017towards}, conducting user behavior analysis in social networks \cite{lopez2014recommendation}, or applying Text Mining techniques to identify similar OER-Documents \cite{duffin2007oer}.
Nevertheless, due to the lack of personalized services like high quality search and recommendation, the popularity of OERs has been limited in most user groups (typically educators or lifelong learners) \cite{sun2018heuristic,ruiz2014semantically,chicaiza2015user,ha2011novel,chicaiza2017recommendation}.

In this paper we address the above mentioned challenges and report on the prototype building of a personalized OER recommender system, which helps lifelong learners 1) to be informed about necessary skills required by their current or future jobs and 2) recommend them OERs to facilitate their progress towards mastering those skills. In this paper, after depicting the current state of the art, we reveal the methods and data we used to build up our skill classifier and OER recommender algorithms. Subsequently, we showcase the validation of our first prototype for two jobs (\emph{Data Scientist} and \emph{Mechanical Engineer}), using semi-structured interviews with domain experts. At the end of the paper we conclude our experiences and suggest further research directions.

\section{\uppercase{State of the Art}}
\label{sec:soa}
\subsection{Matching between Jobs and Skills}
Having access to reliable labour market information on skills and jobs is not easy. Currently, only several governments or inter-governmental organizations (the most prominent actors are the US Government, European Commission or Singapore) attempt to build skill inventories and occupational taxonomies (such as ESCO, ISCO or O*NET). Although these taxonomy building efforts have created a stable basis for basic skill analytics (inter-skill relationships, high level matching to competences and occupations), most of these resources are created and maintained by human experts in several time-consuming steps, which makes them expensive and also susceptible to out-dating \cite{djumalieva2018open}. It is therefore not surprising that more and more commercial and research attempts target new ways to obtain real-time labour market information about skills, using and analysing alternative data sets like job vacancy announcement text, resume text, or social media data. These attempts can be clustered into the following three main categories:

\subsubsection{Semantic-based Methods}
This approach builds on ontologies to reveal and organise components of jobs (e.g. skills, tasks) \cite{sibarani2017ontology,castello2014promoting,khobreh2015ontology}. These methods provide meaningful information for stakeholders (i.e. structure of existing jobs, skills and their relationships), however, their dynamicity is limited, since building and maintaining ontologies to cover a wide range of occupations and skills, are currently done manually (by subject matter experts), which is a very costly and time-consuming exercise \cite{hepp2007possible}.  

\subsubsection{Text Mining and Machine Learning Methods}
 A number of studies analyze online vacancy announcements to classify job components (e.g. skills, tasks) according to existing, static taxonomies (e.g. ESCO). This is done to update taxonomies and provide fresh information about labour markets. Most of these papers try to extract features from the vacancies by applying embedding techniques (e.g. word2vec and doc2vec)\cite{colombo2018applying}, Topic Modeling techniques (e.g. LDA) \cite{colace2019towards,colombo2018applying}, TFIDF \cite{karakatsanis2017data} and use classification techniques such as Logistic Regression, SVM, and Random Forest \cite{boselli2018classifying,boselli2018wolmis} or calculate distance \cite{karakatsanis2017data} to assign job vacancies to the their closest job class. Furthermore, a number of papers are focusing on using Text Mining and clustering techniques to find relationships between skills and jobs, and to calculate similarity measures \cite{djumalieva2018open,wowczko2015skills}. These papers build vectors for skills using embedding techniques, Bag of Words \cite{djumalieva2018open}, and apply clustering techniques such as K-Means \cite{djumalieva2018open} to find the structure of related skills and jobs. Contrary to the Ontology-based systems, given that a powerful model is constructed, these methods can automatically extract the required information form job vacancies. However, the identification of such general models remains challenging.

\subsubsection{Content Analysis}
 Several papers focus on specific job areas, and collect related job vacancies from various sources (e.g. job boards, newspapers). Subsequently, they apply content analysis techniques such as counting the number of skills occurrence and skills co-occurrence in order to provide insights about skills in the investigated job area \cite{verma2019investigation,gardiner2018skill,maer2019skill}. Although, these methods are successful when finding and identifying required skills in a given job area, in most cases, they cannot scale. The reason is that mostly these studies use static lists of jobs and skills in their focus areas, which results in a "tunnel vision" and fail to detect new, emerging job components.  

\subsection{OER Recommendation}
The area of OER recommendation systems has enormous development potential. The available literature on OER based content recommendations to learners is currently limited \cite{chicaiza2017recommendation} and there is no signal that factors related to typical lifelong learning goals (skills, jobs) play any role here. To structure recent developments, we clustered available studies into the following four categories:
\subsubsection{Heuristic Method}
\cite{sun2018heuristic} examines the Cold Start problem \cite{lam2008addressing} in the case of new micro OERs. The paper defines rules, based on recommended sequences of learning objects (e.g. some learning objects should be learnt before others) using an existing ontology and calculates a Violation Degree according to the rules. The more a learning path violates the rules, the higher the Validation Degree is. Subsequently, the system recommends and adds new OERs into users' learning paths, based on minimizing the violation degree.
\subsubsection{Semantic and Ontology Based Methods}
\cite{wan2018learning} builds an ontology for learners, learning objects, and their environments to establish similarity measures between learning objects. This is done in order to update learning objects' properties and provide diverse and adaptive recommendations. Some studies make use of ontologies and open source RDF data to leverage semantic content, and define recommendation algorithms suitable for linked data \cite{chicaiza2017recommendation,ruiz2014semantically,sun2017towards}. Moreover, \cite{chicaiza2015user} tries to define an open linked vocabulary to describe user profiles, in order to facilitate recommendations.

\subsubsection{Social Network Analysis}
\cite{lopez2014recommendation} uses social networks to build graphs of OERs and learners. Therefore, it finds tweets which have valid urls, and builds a graph, based on the co-occurrences of tweets' hashtags. They also build a similar graph with users, based on their \emph{mentions} and \emph{retweet}. Finally, they recognize important and influential hashtags, and use density and centrality measures from the graphs to provide recommendations.

\subsubsection{Machine Learning}
\cite{sun2017towards} attempts to classify users (and their demographic features) with the help of Decision Trees and Naive Bayes algorithms to recommend them OERs. Furthermore, \cite{duffin2007oer} uses Document Clustering and LSA in order to find similar OERs and use them for recommendations.

\subsection{Research Question}
Based on the state of the art, it is clear that 1) it is worthwhile and timely to consider labour market information to define learning goals; 2) Efforts to decompose jobs into components suitable for educational purposes are still in their infancy, and 3) the area of OER recommendation systems is an under-researched area, with a number of challenges from a technical (e.g. available algorithms, data integration, scalability) perspective. 
For these reasons the main research questions and objectives of this paper are:
\begin{itemize}
    \item Empower lifelong learners to construct their own learning trajectories on the basis of labour market information and OER based learning content
    \item Create and evaluate a hybrid OER recommendation system prototype, relying on labour market information, learner and OER properties
    \item Create an algorithm to decompose jobs into unique skills and translate those skills into learning objectives
    \item Develop algorithms to match skills (learning objectives) with learning content available in OERs on the basis of learner and OER properties 
    \item Conduct an initial evaluation of our hybrid OER recommendation system prototype against the general project objectives, the applied logic, and its potential contribution to lifelong learning. 
\end{itemize}
 In general, with this work we expect to advance the potential of OERs to handle the increasing need for learning content and instruction \cite{ha2011novel}, through personalized services for learners, based on labor market data.

\section{\uppercase{Methods}}
\label{sec:methods}
In this section, we detail the data and methods we used to identify required skills and their importance levels for jobs, followed by our OER recommendation algorithms. Finally, we illustrate our prototype system, which provides personalized OER recommendations to learners based on individual skill targets.
\subsection{Data Collection}
For the prototyping, we used a crawled sample data-set from Monster.com containing 22,000 job vacancies \footnote{the data-set is accessible from: \url{https://www.kaggle.com/PromptCloudHQ/us-jobs-on-monstercom}}. We used 80\% of our dataset for training and cross validation and 20\% of them as our test set. Moreover, for our OER recommendation, we have used APIs, provided by the following OER providers: SkillsCommons \footnote{\href{https://www.skillscommons.org/}{https://www.skillscommons.org/}} and Wisc-Online \footnote{\href{https://www.wisc-online.com/}{https://www.wisc-online.com/}}.

\subsection{Labour Market Intelligence (LMI)}
\subsubsection{Extracting Skills from Job Vacancies}
Since our aim was to avoid any dependency on existing taxonomies (which are updated slowly), we put existing methods classifying jobs and skills into predefined classes aside, and created a dynamic job-skill matching mechanism to detect skill changes in jobs quickly. As the first step, we constructed a model to find skill related sentences in job vacancies. After an exploratory analysis, we concluded that large number of vacancies do not contain a "Required Skills" section. Therefore, in order to build our model, we selected vacancies with an explicit "Required Skills" section and run the following preprocessing procedure on each of those vacancies:
\begin{itemize}
    \item Deletion of unimportant characters, punctuations and bullet-points
    \item Removal of irrelevant stop words 
    \item Removal of conjunctions, articles, and prepositions
    \item Sentence Tokenization    
    \item Lowercase Conversion
    \item Lematization
\end{itemize}
Altogether we obtained more than 60,000 sentences with this method. This corpus included both sentences, which were mentioned in a "Required Skills" section (we set their label to 1), and also sentences mentioned in other sections in vacancies (we set their label to 0). As a result, we got around 15,000 sentences related to "Required Skills" and labelled as 1 and around 45,000 sentences not related to "Required Skills" and labelled as 0. Subsequently, we applied embedding techniques on word-level n-grams, and built sentence vectors  with averaging word/n-gram embeddings and using Multinomial Logistic Regression model to minimize the classification error \footnote{We used FastText Library in Python for our classification task \cite{joulin2016bag}}. It should be mentioned that word-level n-gram applies the n-gram concept on character level and find the most common sequences of characters. Therefore, vectors are created for each of the extracted sequence of characters and it helps us build vectors for new words (skills), based on our existing vector for the new word's sequences of characters (e.g. building an initial vector for \emph{Mechatronic} based on existing vectors which are extracted from \emph{Elecronic} and \emph{Mechanic}). Applying our model on the test data-set resulted in the detection of 88.7\% balance-accuracy (including precision and recall) of skill-related sentences. Finally, we used TFIDF weighting to detect skill terms in skill related sentences. It should be mentioned that we used \emph{Minimum Document Frequency} of 3 as cut-off point in order to handle typing errors and remove rare words.

\subsubsection{Calculating Skills' Importance for Jobs}
To calculate the importance of particular skills associated with jobs in a specific geographical location, we calculated the rate of skill occurrence in the previous 6 months at the given job location. After normalizing the rates, we use a simple decay function to compute the new importance score, which combines the previous importance scores and the new rates with more weight on the new rates.

\subsection{Recommending OERs}

\begin{table}
\centering
\vspace{7px}
\caption{User Properties.}\label{tab:user}
\resizebox{.47\textwidth}{!}{
\begin{tabular}{|c|c|c|} \hline
\thead{\textbf{Property}} & \thead{\textbf{Values}} & \thead{\textbf{Note}}\\ \hline
Selected Job&\shortstack{Existing\\Jobs}&\shortstack{selected\\by users}\\ \hline
Skills-Levels&\shortstack{[0..100]\\for\\Skills}&\shortstack{determined\\by users}\\ \hline
\shortstack{Personal\\Information}&\shortstack{Location\\Gender\\Education}&\shortstack{entered \\by users}\\ \hline
Pref\_Resources&\shortstack{[0..100] for\\ Resources}&\shortstack{higher tendency\\$\rightarrow$higher value}\\ \hline
Pref\_Length&[0..100]&\shortstack{\emph{preferred\_long}\\and\\\emph{preferred\_short}}\\ \hline
Pref\_Check&[0..100]&\shortstack{prefer assured\\$\rightarrow$closer to 100}\\ \hline
Pref\_Accessibility&[0..100]&\shortstack{prefer higher\\accessibility\\$\rightarrow$closer to 100}\\ \hline
\end{tabular}
}
\end{table}
\subsubsection{Method for Initializing Learners' Properties}
Table \ref{tab:user} depicts learners' properties in our OER recommender prototype. During the initialization of a new user, we capture known properties entered by users (i.e. Personal Information, Skill Level List, and Selected Job), and also a number of properties without values (i.e. Resource scores, Length scores, Quality scores, and Accessibility scores). To set an initial value for these unknown properties, we sample similar users, based on the known properties and use weighted average (based on similarity) of their properties as initial values for unknown properties. This strategy scaffolds the cold start problem of new users. To sample similar users, we use \eqref{eq:similarity} to compute the similarity between user i and j where the Similarity Effect function for user i and j in property k is calculated as \eqref{eq:similarityweight}.

\begin{equation} \label{eq:similarity}
  similarity(i,j) = \frac{\sum_{k=known\_properties} sim\_effect(i,j,k)}{100}
\end{equation}

\begin{equation} \label{eq:similarityweight}
    sim\_effect(i,j,k)= 
\begin{cases}
    equal\_val(k),& \text{same k for i\&j} \\
    0,              & \text{otherwise}
\end{cases}
\end{equation}

Furthermore, the equality value of property k (equal\_val(k)), showing the effect of variable k on similar behaviour (rating) by users, is calculated through the following process:
\begin{enumerate}
    \item We collect user pairs who gave exactly the same ratings for the same OER in the period
    \item Compute the ratio of the number of pairs having exactly the same value in property k to the number of all pairs
    \item Normalizing the ratios in a way that sum of all the ratios becomes equal to 100 and the normalized ratio of k is the Equality Value of k
\end{enumerate}
This process is executed regularly, after defining a time period (e.g. after every month). 

\subsubsection{Method for Updating Learners' Properties}
Since we aim to capture learners' preferences quickly and provide relevant OERs according to the changes and improvements in learners' property values, we decided to update user properties after each rating action on any of the recommended OERs. This is done by using a real-time updating process that, according to the rating score and the properties of the recommended OERs (i.e. length, quality, accessibility), updates the properties of the users. As a consequence, if a learner is satisfied (dissatisfied) by a given OER, we will encourage (discourage) the properties (see details in the next section) of that particular OER for that learner. For instance, if a user is dissatisfied by a long OER (e.g. it takes 10 week to complete), we will update the \emph{Preferred Long} property of the user and decrease its value in order to provide shorter OERs in the future. Along the same line, with assigning positive ratings to accessible OERs, learners can enhance their accessibility criterion and increase their \emph{Preferred Accessibility} value to receive content with accessibility support (critical for instance for visually impaired learners \cite{elias2017ontology}).

\begin{table}
\centering
\vspace{7px}
\caption{OER Properties.}\label{tab:oer}
\resizebox{.47\textwidth}{!}{
\begin{tabular}{|c|c|c|} \hline
\thead{\textbf{Property}} & \thead{\textbf{Values}} & \thead{\textbf{Note}}\\ \hline
Resource&Repositories&\shortstack{e.g. SkillCommons,\\Wisc-Online}\\ \hline
Skill&\shortstack{Existing\\Skills}&\shortstack{based on subjects}\\ \hline
Author&Full Name&\shortstack{the provider}\\ \hline
URL&URL&\shortstack{web address\\of OERs}\\ \hline
Length&[0..100]&\shortstack{how\_long\\and\\how\_short}\\ \hline
Level&[0..100]&\shortstack{higher level\\$\rightarrow$closer to 100}\\ \hline
Quality&[0..100]&\shortstack{more quality\\assurance\\$\rightarrow$closer to 100}\\ \hline
Accessibility&[0..100]&\shortstack{more accessibility\\$\rightarrow$closer to 100}\\ \hline
Relevance&[0..100]&\shortstack{decreased if\\defined \emph{Irrelevant}}\\ \hline
\end{tabular}
}
\end{table}
\subsubsection{OER Properties}
 Table \ref{tab:oer} shows OER properties. Based on existing literature, we selected \emph{Level}, \emph{Length}, \emph{Quality}, and \emph{Accessibility} as important properties of OERs  \cite{piedra2015seeking,atenas2013quality,elias2018towards}. When assigning a value to a particular OER property, first we extract and order all existing values assigned to that property, then classify them, and count the number of classes. Based on the number of classes, we assign a value between 0 and 100 to that property. For instance, we take property \emph{Level}, we extract 3 values (beginner, intermediate and advanced - 3 classes), and as a result we set the value for beginner OERs to 0, intermediate OERs to 50, and advanced OERs to 100.        
 
\subsubsection{Method for Initializing OER Properties}
For each OER, we attempt to identify similar OERs, based on its known properties. For instance, if we know \emph{Skill} and \emph{Author} of a new OER, we identify all other OERs provided by the same author and the same skill target, compute their average values, and set the initial property values accordingly.
\subsubsection{Method for Updating OER Properties}
Detecting OER properties is a slow process in the beginning, since change happens when users alter their rating pattern. This happens usually when they are confronted with new OERs. Therefore, we run the updating process after a specific time period (e.g. once each month). To adjust the properties (except Relevance) of each OER at first, we collect all related users and their ratings in the given time period. Afterwards, we compute the property values for the OER as X in order to minimize \eqref{eq:loss} using Gradient Descend, where $\theta_i$ is the property vector of user i and $Y_i$ is the satisfaction rate of user i.
\begin{equation} \label{eq:loss}
  Loss Function = \sum_{i=users} |(\theta_i^T * X) - Y_i|
\end{equation}
This strategy of using all recent ratings in updating OER properties, enhances the diversity in our recommendations. All learners contribute to calculating these OER properties (for each OER they studied) through their individual evaluations. Users can also rate OERs as irrelevant. As a consequence, the \emph{Relevance} property of an OER \emph{o} is calculated as \eqref{eq:relevancy} where the \emph{total\_recom(o)} shows the number of times that OER \emph{o} has been recommended to users and \emph{irrelev\_count(o)} is the number of times that \emph{o} has been determined as \emph{Irrelevant}. Finally, OERs with a Relevance Value less than the average in relation to a specific skill, are marked as Irrelevant (for that skill only), and therefore will not be recommended (for that skill) anymore.

\begin{equation} \label{eq:relevancy}
  relevancy(o)=\frac{total\_recom(o) - irrelev\_count(o)}{total\_recom(o)}
\end{equation}

\subsubsection{Recommendation Algorithm}
For recommending an OER to a learner, we calculate Cosine Similarity between the properties of candidate OERs (which are related to the skill-level of any user) and the properties of the user. The system will recommend an OER with the lowest distance between those two. Since we update user properties in a real-time process and update OER properties after a predefined period, for recommending the best match for a user, we only need to find an OER, which has the closest properties to the user. Furthermore, Rating Sparsity problem (i.e. users rate only a few number of OERs) is one of the most important issue when building recommender systems. In our case users and OERs have mutual contribution to calculating properties, which intends to eliminate the effects of Rating Sparsity. Even if an OER has limited amount of ratings, we can rely on the properties of the learners. On the basis of their ratings on other (similar) OERs, we calculate the properties for OERs suffering from Rating Sparsity.  

\subsection{OER Recommender Prototype}
\begin{figure*}
    \centering
    \includegraphics[width=\textwidth]{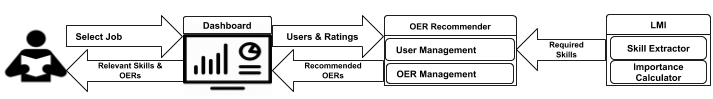}
    \caption{Components of our Labour Market Intelligence (LMI) based OER recommender}
    \label{fig:structure}
\end{figure*}

Learners were confronted with a prototype of our recommender system in a form of a dashboard \footnote{You can find demo of our prototype from: \url{https://github.com/rezatavakoli/CSEDU2020}}. Through this dashboard learners can search for their current or desired job, display the list of required skills and set their level of expertise for each skills. Subsequently, on the learning tab, the dashboard shows the current expertise levels of the learner, and the links to the recommended OERs. OERs are ordered according to the importance of skills for the selected job. In case a learner thinks that a recommended OER is not related (Irrelevant) or does not find the content engaging, a new recommendation could be generated, without changing the expertise level of the learner. After consulting (learning) a recommended OER, learners are asked to rate their satisfaction with that OER. Finally, the dashboard updates the learner's expertise level and provide an updated recommendation based on the new rating. This is done until the learner masters all required skills on the highest level. Figure \ref{fig:structure} depicts the building blocks of our proposed approach.

\section{\uppercase{Validation}}
\label{sec:validation}
To validate our proposed approach, we conducted semi-structured interviews with subject matter experts in the job areas of \emph{Data Science} and \emph{Mechanical Engineering}. We focused on jobs, which are related to these areas, and randomly selected 100 job vacancies for Data Scientists and 100 job vacancies for Mechanical Engineers from August 2019. Afterwards, we applied our skill extraction and importance detection model to select the most important skills in both occupations. To evaluate recommendations, we invited four university instructors with at least 12 years of teaching and 13 years of industrial experience and eight PhD students with a minimum teaching experience of 1 year and a minimum industrial experience of 2 years for a semi-structured interview\footnote{Detailed profiles of our interview participants are available on: \url{https://github.com/rezatavakoli/CSEDU2020}}. Participants gave feedback on our prototype with regards to its general objectives, logic, and potential contribution to individual learning. Each interviewee had to go through the following protocol:
\begin{enumerate}
    \item Learning about the research problem and the proposed approach -  15 minutes
    \item Work with our prototype - 15 minutes
    \item Going through a semi-structured interview with the help of a qualitative questionnaire\footnote{The questionnaire is available on: \url{https://github.com/rezatavakoli/CSEDU2020}} - 30 minutes
\end{enumerate}
During working sessions with our prototype, participants generated more than 150 OER recommendations. 76.9\% of these recommendations were useful and relevant to participants' skill levels and properties. 8.2\% of the recommended OERs were signalled as irrelevant, and in 14.9\% of the cases participants decided to change the recommended OERs. The results of the interviews are summarised in the following three sections.
\subsection{Objectives}
Interviewees confirmed that there is a potential value in building a labour market information driven OER recommender system. Both instructors and PhD students thought that there are several useful and high quality OERs available on the Internet, but finding them are complicated and time-consuming. Regarding the skill extraction, participants recommended to use alternative data sources, besides vacancy announcements. \emph{Student\_2} for example suggested that \say{you should also use other data sources related to labor market like CVs and available data about salaries}. Moreover, interviewees thought that this approach is extremely useful for job-seekers, job-holders, and people who have clear ideas about their preferred occupation. However, they were skeptical about those learners, who want to focus their attention on a specific skill only.
\subsection{Logic}
Participants confirmed that our method to calculate the importance of particular skills in recent job vacancies can potentially help learners to focus on the most important elements of their current or future job. However, as it was also suggested by \emph{Student\_1}, a more intelligent decay function, to combine recent and previous skill important values might be desirable. Regarding the self assessment of learners to set their initial level of expertise, \emph{Instructor\_1} suggested to \say{introduce basic assessment in a form of technical or non-technical questions} for each targeted skill.

\subsection{Contribution to Learning}
Participants emphasized that interacting with learners in order to recognize their preferences (e.g. recommending OERs based on their previous ratings) is one of the most important, novel and engaging component of our proposed approach. \emph{Student\_5} recommended to include more properties: \say{You should capture more learners' properties such as language preferences or type of OERs (e.g. presentation, video).}  Moreover, interviewees were convinced that setting specific and personalized goals for each skill in our prototype system has a strong and positive effect on the learning process.

\section{\uppercase{Conclusion and Future Work}}
\label{sec:conclusion}
In this paper, we showcased a hybrid OER Recommender system prototype to support individual skill development, targeting concrete, labour market oriented skills and jobs. For this prototype a skill extraction mechanism has been constructed, which captures skill related sentences in vacancy announcements with balanced accuracy of 88.7\%. These dynamically generated skills became individual learning objectives and were connected to OER based learning contents. Recommendations were generated through a dashboard, with combining OER and learner properties. The system prototype was validated with semi-structured interviews. The initial results showed that our proposed approach has the potential to aid lifelong learners to construct their individual learning pathways and progress towards their desired job related skills. Moreover, participants valued that user properties were critical, when formulating recommendations. 

We consider this study as an important first step (and a promising positive feedback) on our ongoing research project to empower lifelong learners on the basis of accurate labour market information. We believe that by confronting learners with labour market information, we also support them to develop critical transferable skills such as the awareness of their own learning needs, continuous reflection on their individual learning goals, capacities to (re)design personalised curricula, or measurement of learning achievements. Of course this prototype comes with a number of limitations (e.g. only two jobs were covered; content was only received from two OER repositories, the number of properties for the recommendation were limited), but we believe it is worthwhile to invest further effort in this area. As the next step we plan to expand the context of our investigation with adding more OER repositories to our system, together with extracting more properties from users to provide better recommendations. Moreover, accurate skill decomposition is another key problem to improve, in order to get better assessment about users' expertise level, and to construct more suitable learning pathways for lifelong learners. Finally, we plan to use (quasi-)experimental designs for further developing and validating our prototype in a number of use cases.

\bibliographystyle{apalike}
\bibliography{example}

\end{document}